\documentclass[aip,amsmath,amssymb,reprint]{revtex4-1}

\usepackage{graphicx}
\usepackage[utf8]{inputenc}
\usepackage[T1]{fontenc}
\usepackage{mathptmx}
\usepackage{xcolor}
\usepackage{multibib}
\usepackage[normalem]{ulem}  
\usepackage[colorlinks=true,linkcolor=blue,citecolor=blue,urlcolor=blue]{hyperref}%

\newcites{SM}{SM References}

\newcommand\commentout[1]{}

\begin{document}


\title{Scanning probe-induced thermoelectrics in a quantum point contact}



\author{Geneviève Fleury}
\author{Cosimo Gorini}
\affiliation{Universit\'e Paris-Saclay, CEA, CNRS, SPEC, 91191, Gif-sur-Yvette, France}
\author{Rafael S\'anchez}
\affiliation{Departamento de F\'isica Te\'orica de la Materia Condensada, Condensed Matter Physics Center (IFIMAC), and Instituto Nicol\'as Cabrera, Universidad Autonoma de Madrid, 28049 Madrid, Spain}


\date[]{\href{https://doi.org/10.1063/5.0059220}{Appl. Phys. Lett. {\bf 119}, 043101 (2021)}}

\begin{abstract}
We study three-terminal thermoelectric transport in a two-dimensional Quantum Point Contact (QPC) connected to left and right electronic reservoirs, as well as a third one represented by a scanning probe tip.
The latter acts as a voltage probe exchanging heat with the system but no charges on average. The thermoelectric coefficients are calculated numerically within the Landauer-Büttiker formalism in the low-temperature and linear response regimes. 
We find tip-induced oscillations of the local and non-local thermopowers and study their dependence on the QPC opening. If the latter is tuned on a conductance plateau, the system behaves as a perfect thermoelectric diode: for some tip positions the charge current through the QPC, driven by a local Seebeck effect, can flow in one direction only. 
\end{abstract}


\maketitle 

\begin{figure}[b]
    \includegraphics[keepaspectratio,width=\columnwidth]{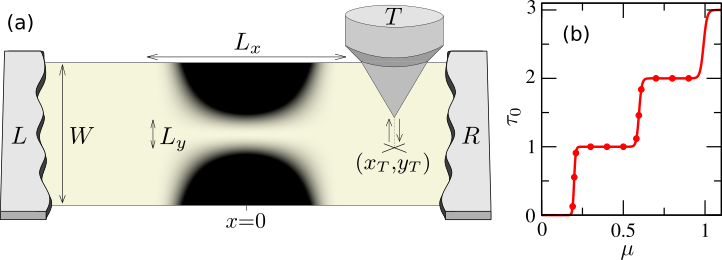}
    \caption{(a) Sketch of our system. A ribbon of width $W$ is attached to two left ($L$) and right ($R$) electronic reservoirs. The onsite confining potential $V_{qpc}(x,y)$ of characteristic length $L_x$ and spatial opening $L_y$ is shown in grayscale. A scanning tip at position $(x_T,y_T)$ is attached to a third electronic reservoir ($T$).
    (b) QPC transmission $\tau_0$ without tip, as a function of $\mu$, in the limit of large $W$. 
    The red dots indicate the values of $\mu$ considered in the top panels of Fig.~\ref{fig_SLT}.
    }
    \label{fig_sys}
\end{figure}

Progress in scanning probe techniques offers new opportunities for gaining understanding of energy transfers at the nanoscale. A major breakthrough was recently achieved in the field with the realization of high-resolution scanning probe thermometers,\cite{Menges2016,Halbertal2016} which allow mapping dissipation in quantum devices. Subsequently, this possibility of measuring local temperature on the nanoscale was leveraged to image the Peltier effect in graphene nanoconstrictions\cite{Harzheim2018} and nanowire heterostructures.\cite{Gachter2020} 
Spatially-resolved images of the Seebeck effect were measured as well by engineering local heaters with scanning tunneling\cite{Park2013} or thermal\cite{Harzheim2018,Gachter2020} microscopes, focused laser~\cite{zolotavin2017} or electron\cite{fast2020} beams, or Joule-heated nanowires.\cite{Mitra2020} 
Contrary to conventional (longitudinal) thermoelectric measurements that are performed across two terminals, such experiments involve a third terminal (\textit{e.g.} the tip of the microscope) and give access to non-local thermoelectric effects.\\
\indent Three-terminal thermoelectrics
has attracted growing interest for a decade,\cite{Benenti2017} leading recently to experimental implementations.\cite{Harzheim2018,Gachter2020,Park2013,zolotavin2017,Mitra2020,roche2015,thierschmann2015,Hartmann2015,jaliel2019,dorsch2020} The prototypical system in this context consists of a central scattering region attached to two (say left and right) electronic reservoirs, and to a third one with which only heat can be exchanged. The third terminal may be a bosonic reservoir,\cite{Rutten2009,Entin-Wohlman2010,Ruokola2012,sothmann2012,Jiang2012,bosisio2016,dorsch2020} or a reservoir of electrons\cite{Sanchez2011,roche2015,thierschmann2015,Hartmann2015} capacitively coupled to the scattering region.
The essentials of such systems can be captured by a simplified model, where the third reservoir acts as a voltage probe,
\textit{i.e.} an electronic reservoir whose electrochemical potential
floats so as to
inject heat but no charge (on average) into the system.\cite{jordan2013} The voltage probe model\cite{buttiker1986} 
is routinely employed to treat
inelastic effects in mesoscopic systems.\cite{Xing2008,Roulleau2009,Kilgour2016,Ma2018} It has also been extensively used in the context of three-terminal thermoelectricity.\cite{DSanchez2011,Mazza2014,Mazza2015,Sanchez2016}
\\
\indent In this paper, we propose non-local and coherent thermoelectric manipulations in a two-dimensional quantum point contact (QPC) via the tip of a scanning tunneling microscope.
The tip acts as a floating third terminal --a movable local voltage probe-- perturbing the phase-coherent electronic propagation through the conductor in a controlled way.\\
\indent QPCs are prototypical mesoscopic systems, whose two-terminal thermoelectric response was the subject of numerous theoretical\cite{streda1989,Proetto1991,Cipiloglu2004,Lunde2005,abbout2011,Whitney2013,Pilgram2015,Kheradsoud2019,Houten1992} and experimental\cite{Houten1992,Molenkamp1992,Dzurak1993,Appleyard1998,brun2019,Yan2019} works.
In particular, the local thermopower of a QPC was recently\cite{brun2019} imaged by scanning gate microscopy (SGM) at low temperature ($25$\,mK). 
In the present paper, the presence of the floating tip turns the basic QPC setup into the three-terminal device sketched in Fig.~\ref{fig_sys}. 
We study numerically the local (longitudinal) and non-local thermoelectric response of the device, for different values of the intrinsic QPC transmission.  The response is characterised by two fundamental features: \textit{(i)} a charge current through the QPC can be induced by a non-local Seebeck effect, \textit{i.~e.~} when the probe is heated, even in configurations where the QPC has no intrinsic thermoelectric response of its own; \textit{(ii)} the probe can rectify a charge current driven by the local Seebeck effect, \textit{i.~e.~} by heating the left (or right) terminal.  We emphasise that both features are extrinsic, non-local and quantum interference-driven, as they require a third terminal and phase-coherent electronic propagation.

Our model is as follows. We introduce first a translation-invariant ribbon of width $W$ discretized on a square lattice (with lattice parameter $a$) and modeling the two-dimensional electron gas (2DEG) in the limit of large $W$. Its tight-binding Hamiltonian reads
\begin{equation}
\label{eq_H2DEG}
H_{2DEG}=-t\sum_{\left\langle i,j\right\rangle}c_i^\dagger c_j + 4t\sum_i c_i^\dagger c_i 
\end{equation}
where 
$c_i^\dagger$ creates a (spinless) electron on site $i$ at position $r_i=(x_i,y_i)$, the sum $\sum_{\left\langle i,j\right\rangle}$ is restricted to nearest neighbors, and $t$ is the hopping parameter. The uniform potential $4t$ in Eq.~\eqref{eq_H2DEG} is included to set the bottom of the ribbon conduction band at zero energy.
We model the QPC with a smooth (symmetric) confining onsite potential $V_{qpc}$ defined by
\begin{equation}
\label{eq_Vqpc}
V_{qpc}(x,y) = \left(\frac{y}{L_y}\right)^2\left[ 1-3\left(\frac{2x}{L_x}\right)^2+2\left|\frac{2x}{L_x}\right|^3\right]^2
\end{equation}
if $|x|<L_x/2$ and $V_{qpc}(x,y)=0$ elsewhere [$L_x$ and $L_y$  controlling respectively the length and the width of the QPC, as illustrated in Fig.~\ref{fig_sys}(a)]. This defines the QPC Hamiltonian $H_{QPC}=\sum_i V_{qpc}(r_i)c_i^\dagger c_i$. Finally, the tip is modeled as a  semi-infinite one-dimensional chain (with lattice parameter $a$) directed along the axis $z>0$ perpendicular to the 2DEG. Its Hamiltonian reads
\begin{equation}
H_{tip}=-t\sum_{\left\langle i,j\right\rangle}c_i^\dagger c_j + 2t\sum_i c_i^\dagger c_i 
\end{equation}
where $c_i$ and $c_i^\dagger$ now operate along $z>0$.
The site $A$ at the extremity of the tip (of coordinates $(x_T,y_T,a)$) is coupled with the hopping term $t_T$ to a single site $B$ of the 2DEG [located at $(x_T,y_T,0)$] below the tip. We denote by $H_c=-t_Tc^\dagger_Ac_B+{\rm h.c.}$ the tunneling Hamiltonian between the tip and the ribbon. 

In the following, we investigate three-terminal thermoelectric transport through the system defined by the Hamiltonian $H=H_{2DEG}+H_{QPC}+H_{tip}+H_c$. The three electronic reservoirs $L$, $R$, and $T$ are formally attached respectively at the left and right extremities of the ribbon and at the end of the tip. They are kept at temperatures $\theta_\alpha$ and electrochemical potentials $\mu_\alpha$ ($\alpha=L$, $R$, or $T$). Within the Landauer-B\"uttiker formalism, the average charge ($I^e_\alpha$) and heat ($I^h_\alpha$) currents flowing from the reservoir $\alpha$ to the scattering region are given by
\begin{subequations}
\label{eqs_I_LB}
\begin{align}
    I^e_\alpha & = \frac{e}{h}\sum_{\beta\neq\alpha}\int\!\mathrm{d}E\tau_{\alpha\beta}(E)[f_\alpha(E)-f_{\beta}(E)] \\
    I^h_\alpha & = \frac{1}{h}\sum_{\beta\neq\alpha}\int\!\mathrm{d}E(E-\mu_\alpha)\tau_{\alpha\beta}(E)[f_\alpha(E)-f_{\beta}(E)]
\end{align}
\end{subequations}
where the sum runs over all reservoirs $\beta\neq\alpha$, $f_\alpha(E)=\{1{+}\exp[(E{-}\mu_\alpha)/k_B\theta_\alpha]\}^{-1}$ is the Fermi-Dirac distribution, $e$ is the electron charge, $h$ and $k_B$ the Planck and Boltzmann constants, and $\tau_{\alpha\beta}(E)$ is the probability for an electron to be transmitted from $\beta$ to $\alpha$ at energy $E$. Hereafter, we take the right reservoir as the reference and set $\mu_R=\mu$, $\theta_R=\theta$. We define $\Delta\mu_\alpha=\mu_\alpha-\mu$, $\Delta \theta_\alpha=\theta_\alpha-\theta$ for $\alpha=L$ and $T$, and assume that
$\Delta\mu_\alpha$ and $\Delta \theta_\alpha$ are small enough
so as to be in the linear response regime.
By expanding the dimensionless currents $\bar{I}^e_\alpha=I^e_\alpha/(ek_B\theta/h)$ and $\bar{I}^h_\alpha=I^h_\alpha/(k_B^2\theta^2/h)$ to first order in $\Delta\mu_\alpha/k_B\theta$ and $\Delta \theta_\alpha /\theta$, one obtains (for $\sigma=e,h$)
\begin{align}
\label{eq:linrespI}
    \bar{I}^\sigma_\alpha & = \sum_{\beta=L,T}\left(L_{\alpha\beta}^{\sigma e}\frac{\Delta\mu_\beta}{k_B\theta}+L_{\alpha\beta}^{\sigma h}\frac{\Delta\theta_\beta}{\theta}\right)
\end{align}
in terms of the Onsager coefficients  $L_{\alpha\beta}^{\sigma\sigma'}$.
Time reversal symmetry implies $L_{\alpha\beta}^{\sigma\sigma'}=L_{\beta\alpha}^{\sigma'\sigma}$ and in the low temperature limit ($\theta\to 0$, to leading order in the Sommerfeld expansion),\cite{Benenti2017} 
\begin{subequations}
\label{eq_sommerfeld}
\begin{align}
    L_{\alpha\beta}^{ee} &= -\tau_{\alpha\beta}(E=\mu) & \mathrm{if}~\alpha\neq\beta\\
    L_{\alpha\beta}^{hh} &= -\tfrac{\pi^2}{3}\tau_{\alpha\beta}(E=\mu)&\mathrm{if}~ \alpha\neq\beta\\
    L_{\alpha\beta}^{eh} = L_{\alpha\beta}^{he} &= -\tfrac{\pi^2}{3}k_B\theta\partial_E\tau_{\alpha\beta}(E=\mu)&\mathrm{if}~ \alpha\neq\beta \\
    L_{\alpha\alpha}^{\sigma\sigma'} &=-\sum_{\beta\neq\alpha}L_{\alpha\beta}^{\sigma\sigma'} &\label{eq_Laa}.
\end{align}
\end{subequations}
In Eq.~\eqref{eq_Laa}, the sum runs over all reservoirs $\beta\neq\alpha$ including 
$R$. Note also that Eq.~\eqref{eq_Laa} resulting from charge and energy conservation is not restricted to low temperatures. \\
\indent The transmission probabilities $\tau_{\alpha\beta}(E)$ and their derivatives $\partial_E\tau_{\alpha\beta}(E)$ are computed using the KWANT software.~\cite{groth2014} Throughout the paper, we take $L_x=100\,a_0$, $L_y=5\,a_0$, $a=0.5\,a_0$, and $t=t_0(a_0/a)^2$, the parameters $a_0\equiv 1$ and $t_0\equiv 1$ defining our space and energy units. 
The choice $a=0.5\,a_0$ allows us to capture the 2DEG continuum limit $a\to 0$ at low energies $0<\mu\lesssim t_0$, yet keeping a tractable computation time. 
Finally, we introduce the notation $\tau_0$ for the QPC transmission $\tau_{RL}(E{=}\mu)$ without tip ($t_T=0)$. With the chosen values of $L_x$ and $L_y$, $\tau_0$ shows well-defined quantized plateaus (at $\tau_0=0,1,2,...$), see Fig.~\ref{fig_sys}(b).

\begin{figure*}[t]
    \includegraphics[keepaspectratio,width=\textwidth]{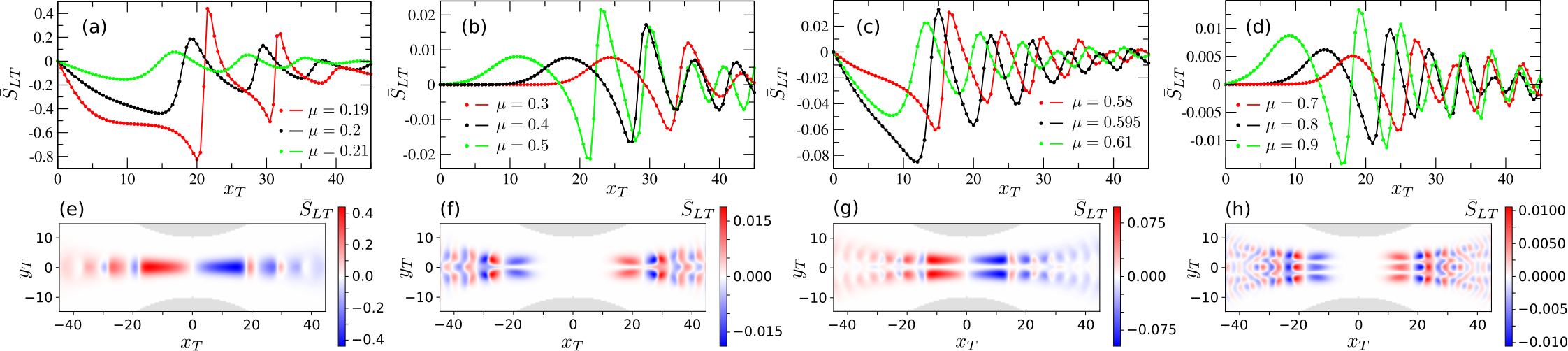}
    \caption{Non-local thermopower $\bar{S}_{LT}$ in the low temperature limit, as a function of the tip position $(x_T,y_T)$, when the QPC is tuned to the first step (first column), first plateau (second column), second step (third column) and second plateau (last column) of conductance.  (Top panels)  $\bar{S}_{LT}(x_T,y_T{=}0)$ [(a) and (d)] and $\bar{S}_{LT}(x_T,y_T{=}3)$ [(b) and (c)] for different values of $\mu$ (indicated by red dots in Fig.\ref{fig_sys}(b)).
    In all panels, data are shown for 
    $W=200$ (full lines) and $W=2000$ (dots).
    (Bottom panels) $\bar{S}_{LT}(x_T,y_T)$ for $\mu=0.2$ (e), $0.4$ (f), $0.595$ (g), and $0.8$ (h) with $W=200$. Data on the grid are locally smoothed for better visibility. Regions where $V_{qpc}(x_T,y_T)\geq 4$ are shown in gray. In all panels, $t_T=0.1t$.}
    \label{fig_SLT}
\end{figure*}


We assume in what follows that the tip acts as a voltage probe, \textit{i.e.} $\Delta\mu_T$ adjusts in such a way that $I^e_T=0$. Due to the Seebeck effect, a temperature bias $\Delta \theta_\alpha$ generates a finite contribution to the charge current $I^e_L=-I^e_R$ through the 2DEG that can be cancelled out by tuning the value of $\Delta\mu_L$. This value is determined by the thermopower\cite{Mazza2014} $S_{L\alpha}=-\Delta\mu_L/(e\Delta \theta_\alpha)$ that has to be calculated under the conditions $I^e_L=I^e_T=0$ and $\Delta \theta_{\alpha'}=0$ for $\alpha'\neq\alpha$. From Eq.~\eqref{eq:linrespI}, we get
\begin{equation}
    \label{eq_SLT1}
    S_{L\alpha}=\frac{k_B}{e}\frac{L^{ee}_{TT}L^{eh}_{L\alpha}-L^{ee}_{LT}L^{eh}_{\alpha T}}{L^{ee}_{LL}L^{ee}_{TT}-L^{ee}_{LT}L^{ee}_{LT}}.
\end{equation}

A non-local response appears when the left and right reservoirs are in equilibrium at the same $\theta$ ($\Delta \theta_L=0$), while the tip reservoir is not ($\Delta\theta_T>0$). 
At low temperatures, it gives
\begin{equation}
    \label{eq_SLT2}
    S_{LT}=\frac{\pi^2}{3}\frac{k_B}{e}k_B\theta\frac{\tau_{LT}\partial_E\tau_{TR}-\tau_{TR}\partial_E\tau_{LT}}{\tau_{LT}\tau_{TR}+\tau_{LR}\tau_{LT}+\tau_{LR}\tau_{TR}},
\end{equation}
where the transmissions and their derivatives are evaluated at $\mu$. We study how $S_{LT}$ depends on the position $(x_T,y_T)$ of the tip, for different values of $\mu$ corresponding to different values of the QPC transmission $\tau_0$. We introduce $\bar{S}_{LT}=S_{LT}/(\pi^2 k_B^2\theta/3e)$ which does not depend on the temperature $\theta$ (as long as the Sommerfeld expansion \eqref{eq_SLT2} is valid\footnote{This is valid
when $k_B\theta$ is much smaller than the typical energy scale $\Delta E$ associated to the oscillations of $\tau_{\alpha\beta}(E)$. While $\Delta E$ decreases with $|x_T|$, the oscillations of $\tau_{\alpha\beta}(E)$ are expected to average out at large $|x_T|$.}).
Our results are summarized in Fig.~\ref{fig_SLT}. Data of $\bar{S}_{LT}$ are shown near the QPC center where finite-width effects in $W$ are negligible and the 2DEG limit is reached. This is illustrated by the superposition of dots (obtained with $W=2000$) on lines (obtained with $W=200$) in the top panels of Fig.~\ref{fig_SLT}.
Note that when $W$ is varied from $200$ to $2000$, the lattice spacing $a$ is kept fixed while the ribbon width is increased.
Moreover, $t_T=0.1t$ was used in Fig.~\ref{fig_SLT} but we have found that $S_{LT}/t_T^2$ is independent of $t_T$ for small $t_T\lesssim 0.2t$ \textit{i.e.} in the limit of a weakly coupled probe (see Supplementary Material).Thus, the choice of the parameters $\theta$, $W$, and $t_T$ in Fig.~\ref{fig_SLT} is irrelevant for the discussion hereafter as long as $\theta\to 0$, $W\gtrsim 200$ and $t_T\lesssim 0.2t$. 

Let us now proceed with the analysis of the oscillations of $S_{LT}(x_T,y_T)$ shown in Fig.~\ref{fig_SLT}. They correspond to fringes of the interferometer formed by the QPC and the tip.\cite{buttiker1989} We find that $S_{LT}(-x_T,-y_T)=-S_{LT}(x_T,y_T)$ and $S_{LT}(x_T,-y_T)=S_{LT}(x_T,y_T)$. This is a direct consequence of the QPC reflection symmetries about the axis $x{=}0$ and $y{=}0$.  In particular, $S_{LT}=0$ when the tip is located at the QPC center $(0,0)$ and preserves 
the QPC spatial symmetries. Moreover, the oscillations decay and eventually vanish when the tip is moved away from the QPC (\textit{i.e.} $S_{LT}\to 0$ when $|x_T|\to\infty$, see Supplementary Material for data at larger $x_T$).
Another important result illustrated in Fig.~\ref{fig_SLT} is the strong dependence on $\mu$ of the non-local thermopower. The amplitude of $S_{LT}$ is much larger on the first QPC transmission step ($0<\tau_0<1$, left column in Fig.~\ref{fig_SLT}) than on higher steps and plateaus ($\tau_0\geq 1$).
\footnote{Formally, $|S_{LT}|$ becomes even larger and larger when the QPC is gradually pinched off ($\tau_0\to 0$) upon decreasing $\mu$, but this regime is out of reach experimentally.} The same behavior is known for the local thermopower $S_{LL}^0=-\left.\Delta\mu_L/(e\Delta \theta_L)\right|_{I^e_L=0}$ of a QPC without tip\cite{Houten1992} and can be understood in the non-local configuration from the analysis of the different terms in Eq.\eqref{eq_SLT2} (see Supplementary Material).

A striking difference between the QPC thermoelectric responses with or without tip appears when $\mu$ is tuned to one transmission plateau. Without tip, electron-hole symmetry around $\mu$ is preserved in the 2DEG, hence $S_{LL}^0=0$ at low temperature: it is not possible to generate a finite charge current ${I^e_L\neq 0}$ through the QPC by Seebeck effect (\textit{i.e.} with $\Delta \theta_L\neq 0$ but $\Delta\mu_L=0$). On the contrary, in the presence of the tip there appear small but finite oscillations around zero of the non-local thermopower $S_{LT}$.  This is because the tip breaks both electron-hole and left-right symmetries (see the second and fourth columns in Fig.~\ref{fig_SLT} corresponding to $\tau_0(\mu)=1$ and $2$ respectively). Thus, a finite charge current ${I^e_L\neq 0}$ can be generated by non-local Seebeck effect (\textit{i.e.} with $\Delta \theta_T\neq 0$ but $\Delta \theta_L= 0$ and $\Delta\mu_L=0$) though the intrinsic thermoelectric response of the QPC vanishes. 

Let us now discuss the colormaps of $S_{LT}(x_T,y_T)$ shown in the lower panels of Fig.~\ref{fig_SLT}. We find a non-trivial dependence on $y_T$ which evolves when $\mu$ is varied : When the QPC is tuned to its first transmission step, $S_{LT}(x_T,y_T)$ has a single-lobe pattern around $y_T=0$ [panel (e)] while two lobes appear near the QPC center on the first plateau [panel (f)] which evolve into two distinct branches on the second QPC step [panel (g)], and eventually a three-lobe pattern emerges on the second plateau [panel (h)]. 
The fact that the number of lobes depends on the opening of the QPC is reminiscent of the behaviour of conductance fringes imaged by SGM,\cite{Topinka2000,Gorini2013} yet this dependence is different in both cases.
Indeed, on the $n$-th QPC plateau, $n+1$ lobes are visible in Figs.~\ref{fig_SLT}(f) and \ref{fig_SLT}(h) while the SGM conductance interference fringes show $n$ lobes.\cite{Topinka2000,Gorini2013}


\begin{figure*}[t]
    \includegraphics[keepaspectratio,width=\textwidth]{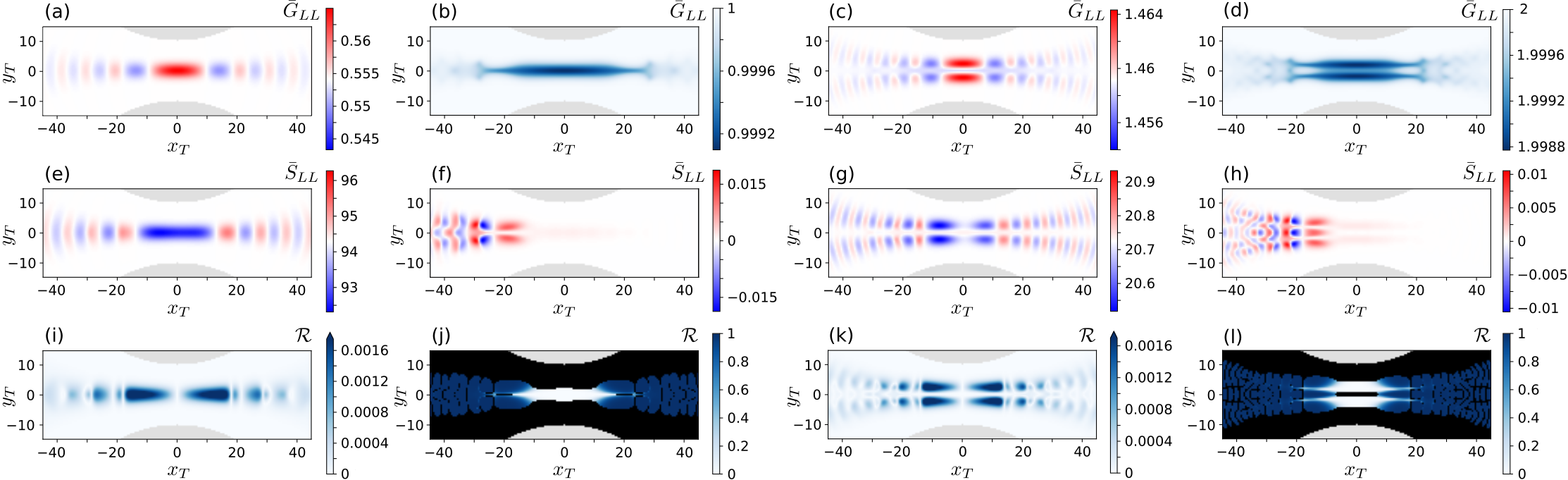}
    \caption{Effective conductance $\bar{G}_{LL}$ (first row), local thermopower $\bar{S}_{LL}$ (second row), and rectification parameter $\mathcal{R}$ (third row) in the low temperature limit, as a function of the tip position $(x_T,y_T)$, for $\mu=0.2$ (first column), $0.4$ (second column), $0.595$ (third column), and $0.8$ (last column). In the top and middle panels, the colormaps are chosen so as the white color corresponds to the values of $\bar{G}^0_{LL}$ and $\bar{S}^0_{LL}$ respectively. In the bottom panels, data for $\mathcal{R}$ with vanishingly small rectified currents ($(|I_\rightarrow|+|I_\leftarrow|)/(\pi^2ek_B^2\theta\Delta\theta/3h)<0.001$) are shown in black. In all panels, $W=200$, $t_T=0.1t$ and regions where $V_{qpc}(x,y)\geq 4$ are shown in gray. 
    }
    \label{fig_rect}
\end{figure*}

We will now explore the longitudinal Seebeck effect and see how it can be leveraged in our system to implement a thermoelectric diode. We assume the temperature bias is now finite in the left lead while $\Delta \theta_T=0$. The tip still plays the role of a voltage probe \textit{i.e.} $\Delta\mu_T$ is determined by imposing $I^e_T=0$. However, $I^h_T\neq 0$ in general. In that configuration, the linear response charge current flowing through the QPC reduces to 
\begin{equation}
    \label{eq_IeL_diode}
    I^e_L=G_{LL}\left(\frac{\Delta\mu_L}{e}+S_{LL}\Delta \theta_L\right),
\end{equation}
where 
the effective two-terminal conductance is given by
\begin{equation}
    G_{LL}=\frac{e^2}{h}\left[ L^{ee}_{LL}-\frac{(L^{ee}_{LT})^2}{L^{ee}_{TT}}\right],
\end{equation}
and the local thermopower $S_{LL}$ is obtained by taking $\alpha=L$ in Eq.~\eqref{eq_SLT1}.
At low temperatures:
\begin{align}
    \label{eq_GLL2}
    G_{LL}&= \frac{e^2}{h}\left[\tau_{LR}+\frac{\tau_{LT}\tau_{TR}}{\tau_{LT}+\tau_{TR}}\right]\\
    \label{eq_SLL2}
    S_{LL}&=\frac{\pi^2}{3}\frac{k_B}{e}k_B\theta\frac{\tau_{LT}\partial_E\tau_{LR}+\tau_{TR}\partial_E\tau_{LT}+\tau_{TR}\partial_E\tau_{LR}}{\tau_{LT}\tau_{TR}+\tau_{LR}\tau_{LT}+\tau_{LR}\tau_{TR}}\,.
\end{align}
The (dimensionless) transport coefficients $\bar{G}_{LL}=G_{LL}/(e^2/h)$ and $\bar{S}_{LL}=S_{LL}/(\pi^2 k_B^2\theta/3e)$ given by Eqs.~\eqref{eq_GLL2} and \eqref{eq_SLL2} are plotted in Fig.~\ref{fig_rect} as functions of the tip position, for the same (four) values of $\mu$ considered in the lower panels of Fig.~\ref{fig_SLT}. As in Fig.~\ref{fig_SLT}, the 2DEG limit $W\to\infty$ is reached in the investigated region (near the QPC center). 
Also, defining $\bar{G}_{LL}^0$ and $\bar{S}_{LL}^0$ in the absence of the tip [$\bar{G}_{LL}^0=\tau_0$ and $\bar{S}_{LL}^0=(\partial_E{\tau}_0)/\tau_0$ at low temperature], we find that
$(\bar{G}_{LL}-\bar{G}_{LL}^0)/t_T^2$ and $(\bar{S}_{LL}-\bar{S}_{LL}^0)/t_T^2$ are independent of $t_T$ in the weak coupling limit ($t_T\lesssim 0.2t$), see Supplementary Material.
When the QPC transmission is tuned on a step, $\bar{G}_{LL}$ oscillates around $\bar{G}^0_{LL}$ [Figs.~\ref{fig_rect}(a) and (c)]. On the plateaus, the tip-induced corrections to $\bar{G}^0_{LL}$ are of smaller amplitude and always negative [Figs.~\ref{fig_rect}(b) and (d)]. In both cases,\footnote{The convergence is not obvious in  Figs.~\ref{fig_rect}(b) and \ref{fig_rect}(d) limited to $|x_T|<45$ but we have checked it by considering larger values of $x_T$ (and of $W$ to eliminate finite-size effects that arise away from the QPC).} $\bar{G}_{LL}\to\bar{G}^0_{LL}$ when $|x_T|\to\infty$. We notice that the behaviour of $G_{LL}$ described above is similar to the one of the two-terminal conductance of a QPC in a SGM configuration.\cite{Topinka2000,Jura2009,Jalabert2010,Gorini2013} Furthermore, we check that $\bar{G}_{LL}(x_T,y_T)$ is symmetric with respect to the axis $x_T=0$. Indeed, when $(x_T,y_T)\to(-x_T,y_T)$, exchanging indices $L\leftrightarrow R$ in Eq.~\eqref{eq_GLL2} and using $\tau_{\alpha\beta}=\tau_{\beta\alpha}$, $\bar{G}_{LL}$ remains invariant.
On the contrary, $\bar{S}_{LL}$ is asymmetric with respect to $x_T=0$. This can be understood from Eq.~\eqref{eq_SLL2} likewise and stems from the fact that $\tau_{TR}\partial_E\tau_{LT}\neq\tau_{TL}\partial_E\tau_{RT}$ in general (see Supplementary Material for more details). While the asymmetry is small when the QPC is tuned to a transmission step [Figs.~\ref{fig_rect}(e) and (g)], it becomes prominent on the plateaus: then the oscillations are strongly suppressed if the tip and the hot terminal are separated by the QPC [Figs.~\ref{fig_rect}(f) and (h)]. In both cases, $\bar{S}_{LL}$ oscillates around (and converges toward) the intrinsic QPC thermopower $\bar{S}_{LL}^0$, as the tip is moved away from the QPC.

The asymmetry of the local thermopower patterns gives rise to current rectification effects. To make it clear, let us compare for a fixed position of the tip the (forward) current $I_\rightarrow = I^e_L$ when $\theta_L=\theta+\Delta \theta$ and $\theta_R=\theta_T=\theta$, to the (backward) current $I_\leftarrow = I^e_R$ when $\theta_R=\theta+\Delta \theta$ and $\theta_L=\theta_T=\theta$ (with otherwise $\mu_L=\mu_R=\mu$).
We calculate $I_\rightarrow=S_{LL}\,G_{LL}\,\Delta\theta$ from Eqs.~\eqref{eq_GLL2} and \eqref{eq_SLL2}, while $I_\leftarrow$ can be computed as well by reproducing the calculations from Eq.~\eqref{eqs_I_LB} with the left lead $L$ (instead of $R$) as the reference. We check that $I_\leftarrow(x_T,y_T)=I_\rightarrow(-x_T,y_T)$ as imposed by the symmetry of the system. In general, $I_\rightarrow(x_T,y_T) \neq I_\leftarrow(x_T,y_T)$ [since $S_{LL}(x_T,y_T)\neq S_{LL}(-x_T,y_T)$].
To quantify the effect, we introduce the rectification parameter $\mathcal{R}=|I_\rightarrow-I_\leftarrow|/(|I_\rightarrow|+|I_\leftarrow|)$. Obviously, $\mathcal{R}(x_T,y_T) =\mathcal{R}(-x_T,y_T)$ and \footnote{Within linear response, $\mathcal{R}$ can also be written as $\mathcal{R} =|S_{LT}|/(|S_{LL}|+|S_{LT}+S_{LL}|)$. With this formula, the discussion held after Eq.~\eqref{eq_R} can be rephrased as follows. On the QPC steps, $|S_{LL}|\gg |S_{LT}|$ and $\mathcal{R}$ is tiny.
On the QPC plateaus, there are some tip positions for which $S_{LL}(x_T,y_T)=0$ and so $\mathcal{R}=1$. In that case, $S_{LL}(-x_T,y_T)=-S_{LT}(-x_T,y_T)$ (compare Fig.~\ref{fig_SLT}(f) to Fig.~\ref{fig_rect}(f) and Fig.~\ref{fig_SLT}(h) to Fig.~\ref{fig_rect}(h)), hence $\mathcal{R}(-x_T,y_T)=1$ too and we recover the expected symmetry law $\mathcal{R}(x_T,y_T)=\mathcal{R}(-x_T,y_T)$.}
\begin{equation}
    \label{eq_R}
    \mathcal{R}(x_T,y_T) = \frac{|S_{LL}(x_T,y_T)-S_{LL}(-x_T,y_T)|}{|S_{LL}(x_T,y_T)|+|S_{LL}(-x_T,y_T)|}\,.
\end{equation}
When the QPC is tuned to a transmission step, the asymmetry of $S_{LL}$ is weak in comparison to the amplitude of $S_{LL}$ (\textit{i.e.} the numerator in Eq.~\eqref{eq_R} is negligible compared to the denominator) and hence, $\mathcal{R}$ is tiny [$\mathcal{R}\leq 0.002$ in 
Figs.~\ref{fig_rect}(i) and (k)].
On the contrary, the strong asymmetry on the QPC plateaus leads to high rectification coefficients, ${\cal R}\approx1$, and the system behaves as an efficient thermoelectric diode [see Figs.~\ref{fig_rect}(j) and (l)]. 
In other words, for $x_T>0$ sufficiently far from the QPC, it is possible to generate a finite $I_\leftarrow$ by heating up the right reservoir, but not to generate a current $I_\rightarrow$ by heating up the left one (or conversely if $x_T<0$). 
The rectification effect is perfect, however the generated currents $\propto \bar{S}_{LL}\bar{G}_{LL}$ are small at the plateaus (compared to those generated at the steps), but in principle measurable.
Note that other thermoelectric diodes were studied in the literature in various contexts.\cite{Kuo2010,Matthews2012,Zhang2012,rossello2017,Craven2018} In our case, the optimal rectification effect exists within linear response 
and results from the 
simultaneous interference-induced broken left-right symmetry and the fact that the tip is allowed to exchange heat with the 2DEG (if $I^h_T=0$ is imposed, $\mathcal{R}=0$).
We note also that a dual rectification effect of the heat current through the QPC is obtained if a voltage bias with respect to $\mu$ is applied between the left and right leads (instead of a temperature bias) and the tip plays the role of a thermal probe (with $I^h_T=0$ and $\mu_T=\mu$) instead of a voltage probe. Within linear response and up to the lowest order of the Sommerfeld expansion \eqref{eq_sommerfeld}, both rectification effects are controlled by the same rectification parameter given by Eqs.~\eqref{eq_R} and \eqref{eq_SLL2}.

In conclusion, we studied low-temperature non-local thermoelectric effects in a QPC coupled to a scanning voltage probe, in linear response to small thermal/voltage biases. The probe is floating, \textit{i.~e.~}it exchanges heat but no (average) charge with the 2DEG.  However it affects the thermoelectric response of the three-terminal device, by two main features.  First, a finite non-local thermopower oscillates as a function of the probe position, even when the QPC is open on a conductance plateau -- and thus its intrinsic (without probe) low-temperature thermopower vanishes.  The oscillations are signatures of the electronic interferometer formed by the QPC and the probe: When the latter is heated, it injects (neutral) electron-hole excitations into the 2DEG, which induce a net charge current through the QPC as the interferometer breaks both left-right and electron-hole symmetries.
Second, the local thermopower oscillates as a function of the probe position as well, around its intrinsic value.  When the QPC is open on a conductance plateau, such oscillations are visible if the tip is on the hot reservoir side, but quickly die out otherwise.  This asymmetric quantum interference pattern leads to potentially perfect current rectification.  Such a rectification is enabled by the presence of the tip, thus bypassing the fundamental limitations of standard (two-terminal) linear response theory.

Our results are proof-of-principle and concern small currents in a bare-bone QPC interferometer.  We hope that they will motivate further (experimental) investigations of the thermoelectric response of nanostructures with scanning probe techniques.

\begin{acknowledgments}
G. F. thanks Rodolfo Jalabert and Dietmar Weinmann for useful discussions. R. S. acknowledges funding from the Ram\'on y Cajal program RYC-2016-20778, and the Spanish Ministerio de Ciencia e Innovaci\'on via grant No. PID2019-110125GB-I00 and through the ``Mar\'{i}a de Maeztu'' Programme for Units of Excellence in R{\&}D CEX2018-000805-M.  C. G. acknowledges stimulating discussions with the STherQO members. 
\end{acknowledgments}

\section*{Supplemental Material}
In the Supplementary Material, we show additional data for various $W$ and $t_T$, and additional plots of the transmissions $\tau_{\alpha\beta}$ and their derivatives $\partial_E \tau_{\alpha\beta}$ entering Eqs.\eqref{eq_SLT2} and \eqref{eq_SLL2}. The effect of auxiliary fictitious probes playing the role of (invasive) local thermometers is also investigated.


\section*{Data availability}
The data that support the findings of this study are available from the corresponding author upon reasonable request.

\section*{References}
\bibliography{biblio2}

\clearpage

\pagebreak



\newcommand{\beginsupplement}{%
        \setcounter{table}{0}
        \renewcommand{\thetable}{S\arabic{table}}%
        \setcounter{figure}{0}
        \renewcommand{\thefigure}{S\arabic{figure}}%
     }
     
\beginsupplement

\section*{Scanning probe-induced thermoelectrics in a quantum point contact: Supplementary Material} 




As in the main text, we fix $L_x=100\,a_0$, $L_y=5\,a_0$, $a=0.5\,a_0$, and $t=t_0(a_0/a)^2$, $a_0\equiv 1$ and $t_0\equiv 1$ being our space and energy units. 

\section{Finite-size effects}
\begin{figure}[b]
    \includegraphics[keepaspectratio,width=\columnwidth]{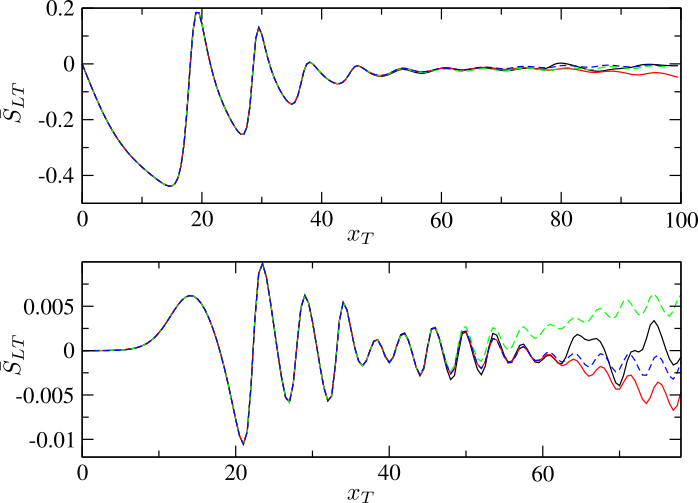}
    \caption{Non local thermopower $\bar{S}_{LT}$ in the low temperature limit as a function of the tip position $x_T$, for various values of the ribbon width $W$ ($W=100$ (black line), $500$ (red line), $1000$ (green dashed line), and $2000$ (blue dashed line)). Data are shown for $\mu=0.2$ (top panel) and $\mu=0.8$ (bottom panel). In both panels, $y_T=0$ and $t_T=0.1t$.}
    \label{fig_SLT_vs_W}
\end{figure}

The low-temperature transport coefficients $\bar{S}_{LT}$, $\bar{S}_{LL}$, and $\bar{G}_{LL}$ are strongly dependent on the ribbon width $W$. As noticed in the main text, the 2DEG limit is guaranteed as long as the tip is moved in a small region around the QPC center (the smaller $W$, the smaller the region). Outside this region, additional scattering against the ribbon boundaries become relevant. We show in Fig.\ref{fig_SLT_vs_W} how the interference patterns of $\bar{S}_{LT}$ vary with $W$. For small $\mu=0.2$ (\textit{i.e.} when the QPC is tuned to its first transmission step, top panel in Fig.~\ref{fig_SLT_vs_W}), finite-width effects along the axis $y_T=0$ are negligible for $|x_T|\lesssim 80$ and $W\geq 100$, and we check that $\bar{S}_{LT}\to 0$ at large $|x_T|$. At larger $\mu=0.8$ (\textit{i.e.} when the QPC is tuned to its second transmission plateau, bottom panel in Fig.~\ref{fig_SLT_vs_W}), stronger finite-width effects in $W$ appear at large $x_T$. They prevent us from providing a rigorous numerical proof of the cancellation of $\bar{S}_{LT}$ away from the QPC, in the 2DEG limit. This would require longer calculations that we did not run.

\begin{figure*}[t]
    \includegraphics[keepaspectratio,width=0.95\textwidth]{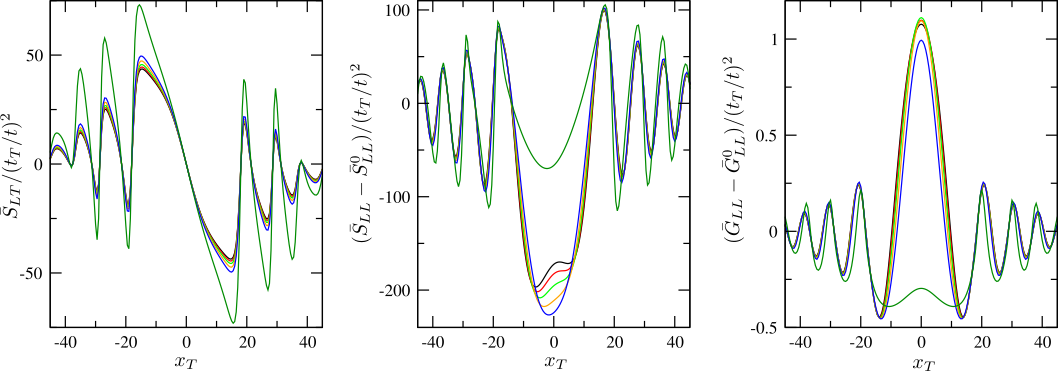}
    \caption{Tip-induced oscillations of the low-temperature transport coefficients ($\bar{S}_{LT}$ (left), $\bar{S}_{LL}$ (middle), and $\bar{G}_{LL}$ (right)), as a function of the tip position $x_T$, for different values of the tip-2DEG hopping term $t_T$ ($t_T=0.01t$ (black line), $0.2t$ (red line), $0.3t$ (green line), $0.4t$ (orange line), $0.5t$ (blue line), and $t$ (dark green line)). After shifting and scaling the data for $\bar{S}_{LT}$, $\bar{S}_{LL}$, and $\bar{G}_{LL}$ along the $y-$axis, the curves of $\bar{S}_{LT}/(t_T/t)^2$, $(\bar{S}_{LL}-\bar{S}_{LL}^0)/(t_T/t)^2$, and $(\bar{G}_{LL}-\bar{G}_{LL}^0)/(t_T/t)^2$ are nearly superimposed for small $t_T\ll t$. In all panels, $y_T=0$, $W=500$ and $\mu=0.2$.}
    \label{fig_all_vs_ttip}
\end{figure*}

\section{Dependence of the transport coefficients on the tip-2DEG coupling}
The amplitude of the tip-induced oscillations of the low-temperature transport coefficients $\bar{S}_{LT}$, $\bar{S}_{LL}$, and $\bar{G}_{LL}$ increases with the tip-2DEG hopping term $t_T$. We find numerically that $\bar{S}_{LT}/(t_T/t)^2$, $(\bar{S}_{LL}-\bar{S}_{LL}^0)/(t_T/t)^2$, and $(\bar{G}_{LL}-\bar{G}_{LL}^0)/(t_T/t)^2$ are (almost) independent of $t_T$ in the low coupling limit ($t_T\ll t$). This is illustrated in Fig.~\ref{fig_all_vs_ttip}.

\section{Transmission plots}
\begin{figure*}[t]
    \includegraphics[keepaspectratio,width=\textwidth]{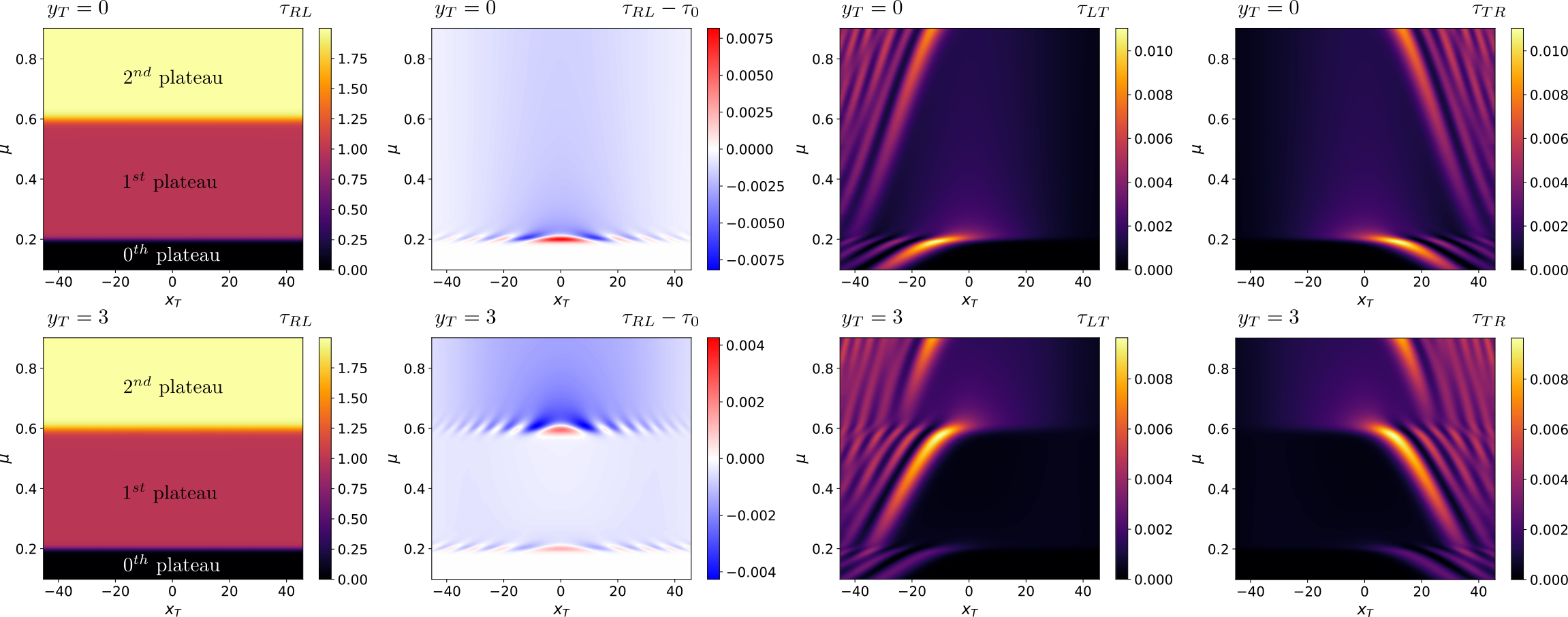}
    \caption{Transmission maps as a function of $\mu$ and $x_T$ for $y_T=0$ (top line) and $y_T=3$ (bottom line). From left to right, colormaps are shown for $\tau_{RL}$, $\tau_{RL}-\tau_0$, $\tau_{LT}$, and $\tau_{TR}$ evaluated at the energy $\mu$. The three regions where $\tau_0(\mu)=0$, $1$ and $2$ (corresponding respectively to the zero-th, first, and second QPC plateau) are clearly visible in the leftmost panels for $\tau_{RL}$ since the deviation of $\tau_{RL}$ from its value without tip $\tau_0$ is small compared to 2 everywhere. In all panels, $W=100$ and $t_T=0.1t$.}
    \label{fig_Tmaps}
\end{figure*}

\begin{figure*}[t]
    \includegraphics[keepaspectratio,width=\textwidth]{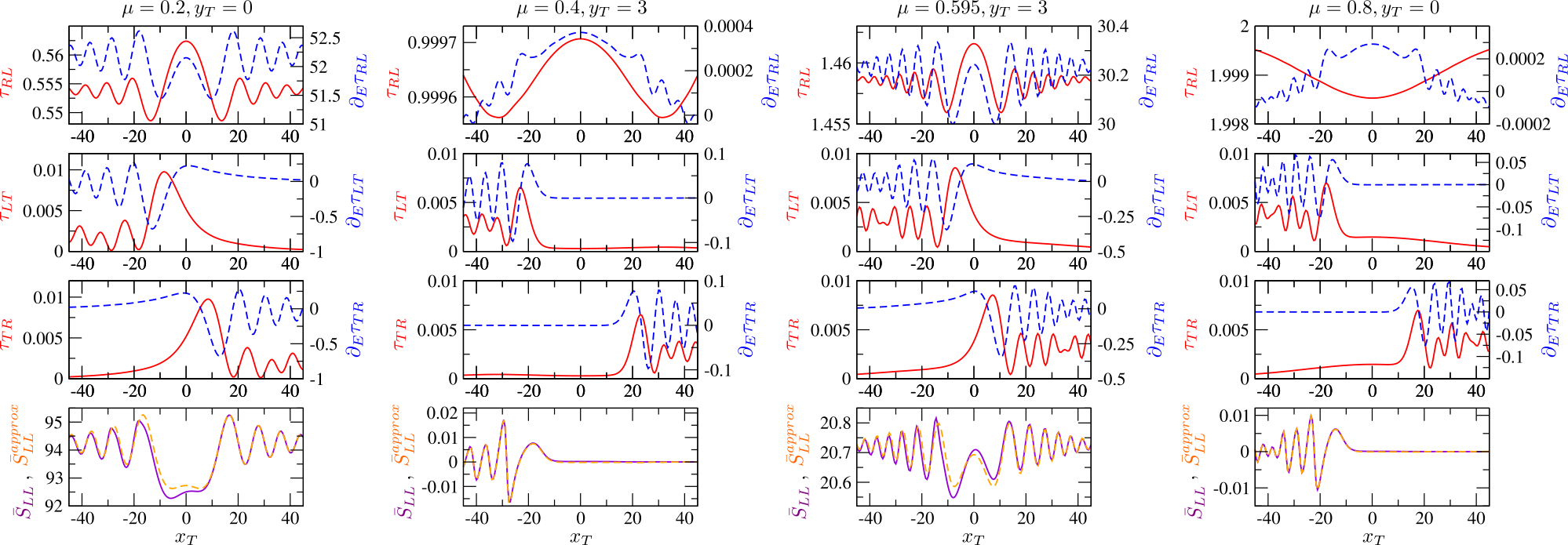}
    \caption{(First line) $\tau_{RL}$ (red full line) and $\partial_E \tau_{RL}$ (blue dashed line) as a function of $x_T$, along $y_T=0$ at $\mu=0.2$ (first column) and $\mu=0.8$ (fourth column), and along $y_T=3$ at $\mu=0.4$ (second column) and $\mu=0.595$ (third column). (Second line) Same for $\tau_{LT}$ (red full line) and $\partial_E \tau_{LT}$ (blue dashed line). (Third line) Same for $\tau_{TR}$ (red full line) and $\partial_E \tau_{TR}$ (blue dashed line). Note that $\tau_{TR}(x_T)=\tau_{LT}(-x_T)$. (Last line) Same for the local thermopower $\bar{S}_{LL}$ (purple full line) and the approximate value $\bar{S}_{LL}^{approx}$ (orange dashed line) defined as $\bar{S}_{LL}^{approx}=\partial_E \tau_{RL}/\tau_{RL}$ for $\mu=0.2$ and $\mu=0.595$ (\textit{i.e.} on the QPC steps) and as $\bar{S}_{LL}^{approx}=\tau_{TR}\partial_E \tau_{LT}/[\tau_0(\tau_{LT}+\tau_{TR})]$ for $\mu=0.4$ and $\mu=0.8$ (\textit{i.e.} on the QPC plateaus). In all panels, $W=100$ and $t_T=0.1t$.}
    \label{fig_T}
\end{figure*}

In the low temperature limit and within linear response regime, the (dimensionless) local and non-local thermopowers $\bar{S}_{LL}$ and $\bar{S}_{LT}$ are controlled by the transmissions $\tau_{RL}$, $\tau_{LT}$, $\tau_{TR}$ and their derivatives $\partial_E \tau_{RL}$, $\partial_E \tau_{LT}$, $\partial_E \tau_{TR}$, through Eqs.\,(8) and (12) of the main paper (with $\tau_{\alpha\beta}=\tau_{\beta\alpha}$ for $\alpha, \beta=L$, $R$ or $T$). To understand the behavior of $\bar{S}_{LL}$ and $\bar{S}_{LT}$ with the tip, it is therefore instructive to investigate how the transmissions $\tau_{\alpha\beta}$ and their derivatives $\partial_E \tau_{\alpha\beta}$ vary when the tip is moved above the 2DEG. Our results are summarized in Figs.\,\ref{fig_Tmaps} and \ref{fig_T}. In those figures, the tip is moved along the $x$ axis at fixed $y_T=0$ or $y_T=3$.\\
\indent In Fig.\,\ref{fig_Tmaps}, we see that for $\mu\lesssim 0.2$ \textit{i.e} when $\tau_0(\mu)=0$ (see Fig.\,1 of the main paper), the left-to-right transmission of the three-terminal device vanishes ($\tau_{RL}=\tau_0=0$) since the QPC is closed. However, if the tip is placed  between the QPC and the left lead, electrons coming from the tip can flow towards the left lead, after a sequence of bounces against the QPC barrier. This explains why interference fringes are visible for $x_T<0$ at small $\mu\lesssim 0.2$ in the colormaps of $\tau_{LT}(x_T,\mu)$ shown in Fig.\,\ref{fig_Tmaps} (third column). For symmetry reasons, the same behavior is observed for $\tau_{TR}(x_T)=\tau_{LT}(-x_T)$ (last column of Fig.\,\ref{fig_Tmaps}). When $\mu$ is increased so as one electronic mode can be transmitted through the QPC ($0<\tau_0(\mu)\leq 1$), $\tau_{RL}$ increases. The latter is almost unaffected by the presence of the tip on the first QPC plateau while deviations of $\tau_{RL}$ from its value $\tau_0$ without tip are relevant around the first QPC step. In all cases, $\tau_{RL}(x_T)=\tau_{RL}(-x_T)$. On the contrary, the colormaps of $\tau_{LT}$ (and $\tau_{TR}$ as well) are strongly asymmetric with respect to the axis $x_T=0$. When $\mu$ is increased further so as $1<\tau_0(\mu)\leq 2$, the second QPC channel comes into play. Due to its V-shaped spatial structure, it has no interplay with the tip if the latter is moved along the axis $y_T=0$ (top panels in Fig.\,\ref{fig_Tmaps}). However, signatures of the opening of the second QPC channel are clearly visible in the different transmission maps along $y_T=3$ (bottom panels in Fig.\,\ref{fig_Tmaps}).\\
\indent In Fig.\,\ref{fig_T} (first three lines), we show horizontal cuts of the colormaps displayed in Fig.\,\ref{fig_Tmaps} for four values of $\mu$ ($0.2$, $0.4$, $0.595$, and $0.8$) considered in Figs. 2 and 3 of the main paper and for which the QPC is tuned respectively to its first step, first plateau, second step, and second plateau of conductance. Data are shown for the transmissions $\tau_{RL}$, $\tau_{LT}$, $\tau_{TR}$ (red lines) and their derivatives $\partial_E \tau_{RL}$, $\partial_E \tau_{LT}$, $\partial_E \tau_{TR}$ (blue dashed lines). In addition, we also show data for the local thermopower $\bar{S}_{LL}$, obtained using Eq.\,(12) of the main paper (purple lines in the bottom panels of Fig.\,\ref{fig_T}). On the QPC steps (first and third columns in Fig.\,\ref{fig_T}), $\tau_{LT}, \tau_{TR}\ll \tau_{RL}$ and $\partial_E\tau_{LT}, \partial_E\tau_{TR}\ll \partial_E\tau_{RL}$, so that the local thermopower $\bar{S}_{LL}$ is well approximated by the formula (orange dashed lines in the first and third bottom panels) \begin{equation}
    \bar{S}_{LL}\approx \frac{\partial_E \tau_{RL}}{\tau_{RL}}~~\mathrm{on~the~QPC~steps}
\end{equation}
that neglects the small asymmetry $\bar{S}_{LL}(x_T)\neq \bar{S}_{LL}(-x_T)$. On the QPC plateaus (second and fourth columns in Fig.\,\ref{fig_T}), $\tau_{RL}\approx \tau_0$ while $\partial_E \tau_{RL} \ll \partial_E\tau_{LT}, \partial_E\tau_{TR}$, and so $\bar{S}_{LL}$ is very well approximated by the formula (orange dashed lines in the second and fourth bottom panels)
\begin{equation}
    \bar{S}_{LL}\approx\frac{\tau_{TR}\partial_E \tau_{LT}}{\tau_0(\tau_{LT}+\tau_{TR})} ~~\mathrm{on~the~QPC~plateaus}.
\end{equation}
As a result, $\bar{S}_{LL}$ is finite only when the tip is on the left side of the QPC (\textit{i.e.} the one attached to the hot reservoir) because $\partial_E \tau_{LT}\approx 0$ for $x_T>0$ while it is finite for $x_T<0$, and $\tau_{TR}$ is also (small but) finite for $x_T<0$. This explains the origin of the asymmetry in the interference patterns of $\bar{S}_{LL}$ (giving rise to the rectification effect discussed in the main paper) and why this asymmetry is more visible on the QPC plateaus than on the QPC steps.

\section{Beyond the coherent regime}
\label{sec_inelastic}
The results described above assume coherent quantum transport through the device, except in the vicinity of the tip playing the role of a scanning voltage probe. We will now investigate (in a phenomenological way) the effects of incoherent scattering on a large scale around the QPC. For that purpose, we introduce in our system fictitious probes mimicking inelastic (electron-electron) scattering. In a region of width $W_p$ and length $L_p$ around the QPC center, we attach to each site $S_p$ in the 2DEG a semi-infinite chain (with lattice parameter $a$) directed along the axis $z<0$ and described by the Hamiltonian
\begin{equation}
H_{p}=-t\sum_{\left\langle i,j\right\rangle}c_i^\dagger c_j + \mu \sum_i  c_i^\dagger c_i \,.
\end{equation}
We denote by $t_p$ the (spatially uniform) hopping term between the site $S_p$ and its nearest neighbor in the probe $p$. 
At the energy $\mu$ around which transport is investigated, the self-energy of a probe is purely imaginary and reads $\Sigma_p(i,j) = -i (t_p^2/t)\delta_{ij}$. 
Each probe $p$ is also attached to an electronic reservoir characterized by its temperature $\theta_p$ and electrochemical potential $\mu_p$. Their values adjust in such a way that the net average charge and heat currents flowing through the probe vanish (\textit{i.e.} $I^e_p=0$, $I^h_p=0$). Contrary to the tip which acts as a voltage probe, those probes do not exchange heat with the 2DEG. Such probes have been used in the literature to model local thermometers.\cite{Bergfield2013,Meair2014,Shastry2015,Ye2016} Here, we use them to investigate the role of inelastic processes upon the thermoelectric response of our device, in the spirit of Refs.~\onlinecite{DSanchez2011,Saito2011,Brandner2013}. 

We proceed as follows. We compute the set of  $\tau_{\alpha\beta}(\mu)$ and $\partial_E\tau_{\alpha\beta}(E=\mu)$ between the $3+M$ reservoirs, $M$ being the number of fictitious probes. We deduce (with Eq.\,(6) of the main paper) the total Onsager matrix (of dimension $[2(M+2)]^2$) in the low temperature limit. Then we write down Eq.\,(5) of the main paper for the particle and heat currents in the $M$ probes (by noticing that the index $\beta$ in Eq.\,(5) now runs over $L$, $T$, and the $M$ probes) and we impose the probe condition \textit{i.e.} $I^e_p=0$, $I^h_p=0$ to deduce the values of $\Delta\mu_p=\mu_p-\mu$ and $\Delta\theta_p=\theta_p-\theta$ ($p=1,...,M$) as a function of $\Delta\mu_L$, $\Delta\mu_T$, $\Delta \theta_L$, and $\Delta \theta_T$. 
In practice, this requires to solve four systems of $2M$ linear equations. 
We insert those values into Eq.\,(5), now written for $\alpha=L$ and $T$, and define thereby an effective $4\times 4$ Onsager matrix $\mathbf{\tilde{L}}$ that relates through Eq.\,(5) the particle and heat currents $(I^e_L,I^h_L,I^e_T,I^h_T)$ to the biases $(\Delta\mu_L, \Delta \theta_L, \Delta\mu_T, \Delta \theta_T)$. In the end, we can reproduce the study done in Sections III and IV of the main paper in the presence of incoherent scattering processes by replacing therein the coherent Onsager matrix $\mathbf{L}$ with $\mathbf{\tilde{L}}$. The resulting $\bar{S}_{LT}$, $\bar{S}_{LL}$, and $\bar{G}_{LL}$ are independent of $\theta$ in the low temperature limit.\\
\begin{figure}[h]
    \includegraphics[keepaspectratio,width=\columnwidth]{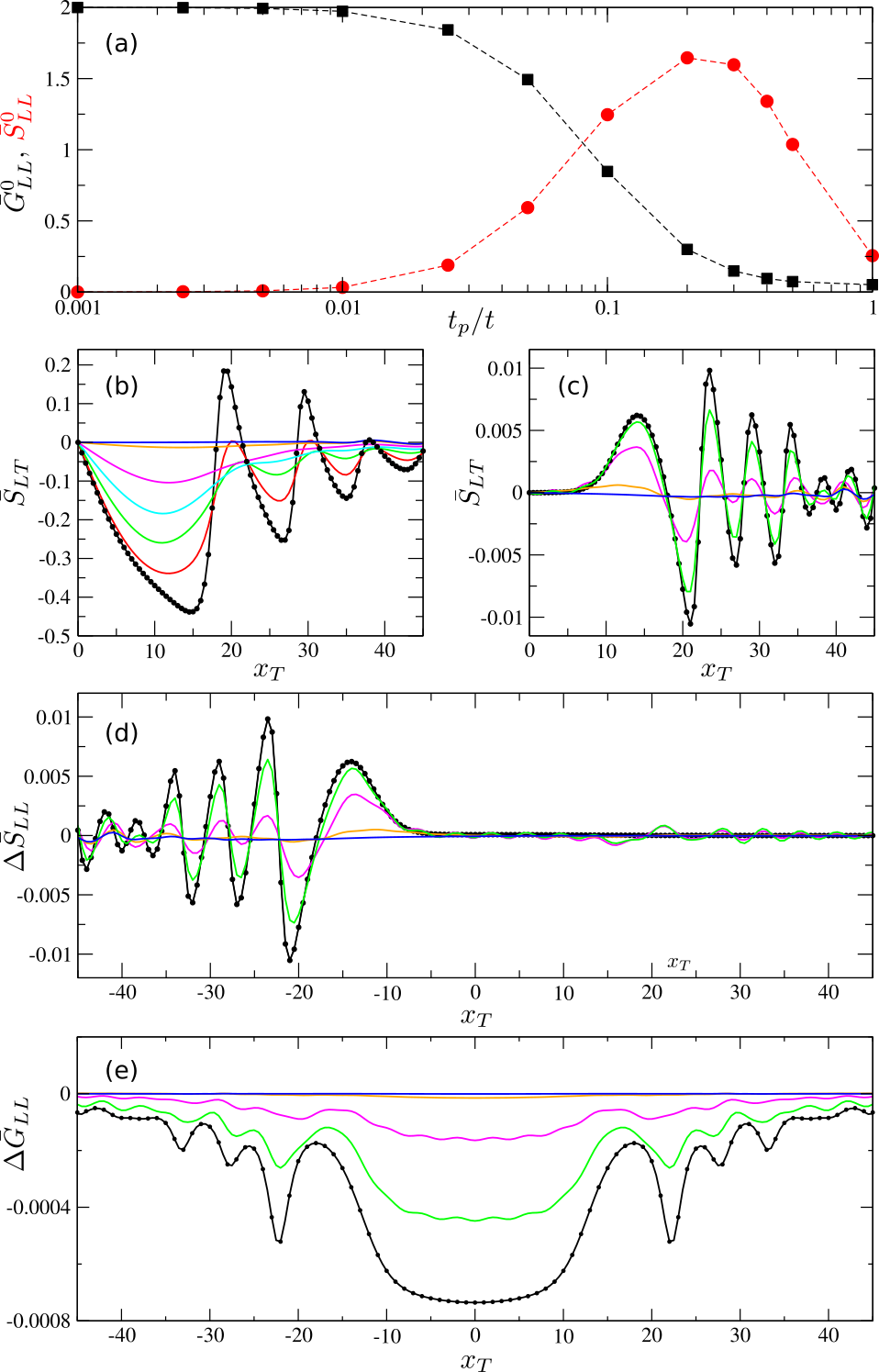}
    \caption{Effect of fictitious probes on the thermoelectric response. (a)~$\bar{G}_{LL}^0$ (black squares) and $\bar{S}_{LL}^0$ (red circles) without tip, as a function of $t_p/t$, for $\mu=0.8$. Dashed lines are guides to the eye. (b) $\bar{S}_{LT}(x_T,y_T=0)$ in the low temperature limit for $\mu=0.2$ and various values of $t_p$ ($t_p=0.001t$ (black line), $0.03t$ (red line), $0.05t$ (green line), $0.07t$ (cyan line), $0.1t$ (pink line), $0.2t$ (orange line), and $0.3t$ (blue line)). (c) Same as (b) for $\mu=0.8$. (d) Same as (c) for $\Delta\bar{S}_{LL}\equiv\bar{S}_{LL}-\bar{S}_{LL}^0$. (e) Same as (d) for $\Delta\bar{G}_{LL}\equiv\bar{G}_{LL}-\bar{G}_{LL}^0$. In panels (b) to (e), data without fictitious probes (corresponding to Figs.2(a), 2(d), 3(d) and 3(h) of the main paper) are shown with black dots. Parameters: $W=100$, $W_p=100$, $L_p=60$, and $t_T=0.1t$ (except in (a) where $t_T=0)$.}
    \label{fig_probes}
\end{figure}
\indent Our results are summarized in Fig.~\ref{fig_probes}. In Fig.\,\ref{fig_probes}(a), we show that the conductance $\bar{G}_{LL}^0$ and the thermopower $\bar{S}_{LL}^0$ of the QPC without tip respectively decreases (till vanishing) and increases (before decreasing and vanishing) with the hopping term $t_p$. Thus, the fictitious probes become invasive as soon as $t_p\gtrsim 0.1t$ : In addition to incoherent scattering processes, they induce backscattering.\cite{Golizadeh2007} In Figs.\,\ref{fig_probes}(b)-(e), we add the tip and explore how the interference patterns of $\bar{S}_{LT}$, $\bar{S}_{LL}$, and $\bar{G}_{LL}$ are modified in the presence of the fictitious probes. We check first that we recover the coherent limit discussed in the main paper when $t_p\to\ 0$ (as evidenced by the superposition of dots on black lines in Figs.\,\ref{fig_probes}(b)-(e)). As long as the fictitious probes are non invasive (\textit{i.e.} in the weak coupling limit $t_p\lesssim 0.01t$), $\bar{S}_{LT}$, $\bar{S}_{LL}$, and $\bar{G}_{LL}$ are unaffected by the probes. For larger $t_p$, the tip-induced oscillations of $\bar{S}_{LT}$, $\bar{S}_{LL}$ and $\bar{G}_{LL}$ -- around $0$, $\bar{S}_{LL}^0$ and $\bar{G}_{LL}^0$ respectively -- die out gradually. The present investigation does not allow us however to disentangle the roles of (invasive) backscattering and (noninvasive) incoherent scattering processes. To assess specifically the effect of quantum coherence on the tip-induced thermoelectric effects discussed in the main paper, other models might be considered.\cite{Golizadeh2007,Chen2012,Forster2005,Rahman2019} This is left for future works.




\end{document}